\documentclass[aps,prd,twocolumn,preprintnumbers,nofootinbib,superscriptaddress,amsmath]{revtex4-1}
\usepackage{graphicx}
\usepackage{amsmath}
\usepackage{amssymb}
\usepackage{amsfonts}
\usepackage{hyperref}
\usepackage{url}
\usepackage{color}
\usepackage{bm}
\usepackage{array}
\usepackage{placeins}
\usepackage{bbold}
\usepackage{algorithm}
\usepackage{algpseudocode}
\usepackage[normalem]{ulem}

\newcommand{\ssout}[1]{}

\def\d{{\rm d}}

\begin{document}
\makeatother

\preprint{IFT-UAM/CSIC-24-064}

\title{Effects of local cosmic inhomogeneities on the gravitational wave event rate}

\author{Gonzalo Morras}
\email{gonzalo.morras@uam.es}
\affiliation{Instituto de F\'isica Te\'orica UAM/CSIC, Universidad Aut\'onoma de Madrid, Cantoblanco 28049 Madrid, Spain}
\author{Juan Garc\'ia-Bellido}
\email{juan.garciabellido@uam.es}
\affiliation{Instituto de F\'isica Te\'orica UAM/CSIC, Universidad Aut\'onoma de Madrid, Cantoblanco 28049 Madrid, Spain}

\date{\today}

\begin{abstract}

The local Universe is highly inhomogeneous and anisotropic. We live in a relatively sparse region of the Laniakea supercluster at the edge of a large 80~Mpc-wide void. We study the effect of these inhomogeneities on the measured gravitational wave event rates. In particular, we estimate how the measured merger rate of compact binaries is biased by the local matter distribution. The effect of the inhomogeneities on the merger rate is suppressed by the low angular resolution of gravitational wave detectors coupled with their smoothly decreasing population-averaged sensitivity with distance. We estimate the effect on the compact binary coalescence event rate to be at most~6\% depending on the chirp mass of the target binary system and the sensitivity and orientation of the detectors.

\end{abstract}
\maketitle

\section{Introduction}
\label{sec:intro}

Detection of gravitational waves (GW) from compact binaries has become routine since their first detection in 2015~\cite{LIGOScientific:2016aoc} by the LIGO~\cite{LIGOScientific:2014pky} and Virgo~\cite{VIRGO:2014yos} collaborations. Since then, around 100 compact binary coalescence (CBC) events have been detected~\cite{KAGRA:2021vkt} in the first three observing runs, and another 80 public alerts in the first part of the fourth observing run~\cite{O4_public_alerts}. Most of the events detected so far correspond to very massive systems at high redshift, where the distribution of matter in the Universe can be well approximated by a homogeneous and isotropic distribution~\cite{Essick:2022slj,Vitale:2022pmu}. 
However, when studying the merger rates of systems with smaller masses such as binary neutron stars (BNS)~\cite{KAGRA:2021duu} or sub-solar mass (SSM) black holes~\cite{LIGOScientific:2022hai}, the distance probed by the detectors is smaller, having BNS ranges smaller than $\sim$200~Mpc for current interferometers~\cite{Abbott_2020}. At distances smaller than $\sim$250~Mpc around Earth, the matter distribution is far from homogeneous, and has to be taken into account.

In fact, we know from local surveys that the Milky Way galaxy lives in a supercluster called Laniakea~\cite{Courtois:2022mxo,Tully:2014gfa,Dupuy:2023ffz}, moving toward the Great Attractor~\cite{Dressler:1986rv} and ''surfing'' a large wall surrounding an 80~Mpc wide void. The local Universe contains huge cosmic structures such as the Sloan Great Wall~\cite{Gott:2003pf}, a filament-shaped over-density located at a distance of $~\sim 200$ Mpc and measuring a staggering $\sim 400$ Mpc, challenging previous assumptions of the homogeneity scale of our Universe~\cite{Alonso:2014xca}. Such an inhomogeneous matter distribution must necessarily have an impact on the probability that a given GW event occurs at a given distance in a given direction, assuming that the binary black hole (BBH) progenitors' location correlates with the matter distribution, i.e., more events toward large concentrations of matter.

The effect of local inhomogeneities may therefore affect our inference of the rate of GW events. By this we mean that we may infer a rate of events that is biased by our location in the Universe, which would have to be taken into account when constraining theoretical models, or when extrapolating the measured rates to estimate rates coming from regions that are farther away than what has been measured. 

In this paper we have studied the effect on the rate of events of the known matter inhomogeneities measured on Ref.~\cite{Courtois:2022mxo} using the peculiar (i.e., gravitational) velocities of galaxies from the CosmicFlows (CF) catalogs~\cite{Tully:2016ppz,Tully:2022rbj}, which compile galaxies in the local Universe whose distances to us have been estimated with different methods independent of Hubble's law. We have used the antenna patterns of the two LIGO detectors (Hanford and Livingston) and the Virgo detector, as well as the rotation of the Earth, which sweeps through the sky and further smears out the sensitivity to cosmic matter angular inhomogeneities. Under these assumptions we find that the effect on the sensitive time-volume $\langle VT\rangle$ for a compact binary coalescence (CBC) search is at most~6\%.

In Sec.~\ref{sec:VT} we study the VT formalism in an inhomogeneous Universe. In Sec.~\ref{sec:VT_estimate} we estimate the effect of the local Universe inhomogeneities in the rate of BBH events and compute the effect of these inhomogeneities for current GW detectors, and in Sec.~\ref{sec:conclusions} we give our conclusions.

A \texttt{python} code to reproduce the results of this paper is publicly available at \url{https://github.com/gmorras/Inhomogeneous_VT}.

\section{The inhomogeneous Universe $\langle VT\rangle$}
\label{sec:VT}

Population analysis and estimation of merger rates of compact binaries rely on the sensitive time-volume ($\langle VT \rangle$) surveyed by the GW searches they are based on. For example, if no event is detected, using loudest event statistics, upper limit constraints on the merger rate can be computed as~\cite{Biswas:2007ni}

\begin{equation}
    R_{\mathrm{C.L.}} = \frac{-\log(1 - C.L.)}{\langle VT \rangle} \, .
    \label{eq:R_constraints}
\end{equation}

\noindent where C.L. indicates the confidence level and it is usually chosen to be $\mathrm{C.L.} = 90\%$. For a given GW search, the sensitive time-volume surveyed can be measured as

\begin{equation}
    \langle V T \rangle \!\! =\!\! \int_{T}\!\! \d t \! \int_V \!\!  \d V_c \frac{1}{1 + z} \frac{R(\vec{x})}{\overline{R}_0} \int \!\!  \d\vec{\theta} \, p_\mathrm{pop}(\vec{\theta}) \epsilon(\vec{x}, t, \vec{\theta}) \, ,
    \label{eq:VT_begin}
\end{equation}

\noindent where the time integral is done in the duration of the search of length $T$, the factor $1/(1+z)$ accounts for the time dilation of the intrinsic rate \cite{LIGOScientific:2016kwr, Tiwari:2017ndi}, $R(\vec{x})$ denotes the local event rate at position $\vec{x}$ and $\overline{R}_0$ is the average rate at redshift $z=0$. 
The rate $R(\vec{x})$ depends on position due to both the rate inhomogeneities induced by the inhomogeneities of the Universe, as well as due to the evolution of the mean merger rate with cosmic time. 
For example, in the case of stellar mass BBHs ($m_\mathrm{BH} > 5 M_\odot$), there is evidence that, for $z \lesssim 1$, the mean merger rate behaves as~\cite{KAGRA:2021duu}

\begin{equation}
    \overline{R}^\mathrm{BBH}(z) = \overline{R}_0^\mathrm{BBH} \cdot (1+z)^\kappa \,  \, \quad \mathrm{with} \;\; \kappa = 2.9^{+1.7}_{-1.8}.
    \label{eq:RBBH_z}
\end{equation}

In Eq.~\eqref{eq:VT_begin}, $\epsilon(\vec{x}, t, \vec{\theta})$ denotes the probability to detect a signal with source position $\vec{x}$ and arrival time $t$ as a function of its intrinsic parameters $\vec{\theta}$, and $p_\mathrm{pop}(\vec{\theta})$ is the distribution function for the astrophysical population. We can define a time and population-averaged probability to detected a signal at position $\vec{x}$ as

\begin{equation}
    \epsilon(\vec{x}) = \frac{1}{T} \int_T \d t  \int \d\vec{\theta} p_\mathrm{pop}(\vec{\theta}) \epsilon(\vec{x}, t, \vec{\theta}) \, .
    \label{eq:eps_avg}
\end{equation}

We also define $\delta_R(\vec{x})$ as the event rate inhomogeneities, i.e. the deviations of the event rate $R(\vec{x})$ from its average at the present epoch $\overline{R}_0$ as 

\begin{equation}
    \delta_R(\vec{x}) = \frac{R(\vec{x}) - \overline{R}_0}{\overline{R}_0} \, ,
    \label{eq:delta_R_def}
\end{equation}

\noindent using Eqs.~(\ref{eq:eps_avg},\ref{eq:delta_R_def}) we can rewrite Eq.~\eqref{eq:VT_begin} as

\begin{equation}
    \langle V T \rangle = T \int_V  \d V_c (1 + \delta_R(\vec{x})) \epsilon(\vec{x}) \frac{1}{1 + z} \, ,
    \label{eq:VT_eps_avg}
\end{equation}

The average probability to detect a signal $\epsilon(\vec{x})$ will depend on its position due to two effects. First of all it will depend on redshift $z$, since the amplitude of the GWs is inversely proportional to the luminosity distance ($h \propto 1/d_L(z)$), making it harder to detect them at farther distances. It will also depend on the position of the source in the sky, since even though GW detectors usually have broad angular sensitivity, they do not have the same sensitivity in all directions~\cite{Chen:2016luc}. For earth-bound detectors, if the duration of the search is much longer than a day ($T\gg 1 \, \mathrm{day}$), we expect the rotation of the earth to average out the dependence of the sensitivity with right ascension $\alpha$, and therefore to have 

\begin{equation}
    \epsilon(\vec{x}) = \epsilon(z, \delta) \, ,
    \label{eq:eps_avg_dependence}
\end{equation}

\noindent where $\delta$ is the declination of the GW source. In the integral of Eq.~\eqref{eq:VT_eps_avg}, the comoving volume element is given by:
\begin{equation}
    \d V_c = d_M^2 \,\d d_c \, \d\alpha \, \d\sin{\delta}
    \label{eq:dVc_dc}
\end{equation}

\noindent where $d_c$ is the comoving distance, defined as 
\begin{equation}
    d_c(z) = \frac{c}{H_0} \int_0^z \frac{\d z'}{E(z')} \, ,
    \label{eq:dc_def}
\end{equation}

\noindent $E(z)$ is the dimensionless Hubble parameter, defined as

\begin{align}
    & E(z) = \frac{H(z)}{H_0} \nonumber \\
    & = \sqrt{\Omega_r (1+z)^4 + \Omega_m (1+z)^3 + \Omega_k (1+z)^2 + \Omega_\Lambda} \, ,
    \label{eq:Ez_def}    
\end{align}

\noindent and in Eq.~\eqref{eq:dVc_dc}, $d_M(z)$ is the comoving transverse distance, defined as 

\begin{equation}
    d_M (z) = 
    \begin{cases}
       \frac{1}{\sqrt{\Omega_k}} \frac{c}{H_0} \sinh\left(\sqrt{\Omega_k} \frac{H_0 d_c(z)}{c} \right) &\; \Omega_k > 0\\
       d_c(z) &\; \Omega_k = 0\\ 
       \frac{1}{\sqrt{|\Omega_k|}} \frac{c}{H_0} \sin\left(\sqrt{|\Omega_k|} \frac{H_0 d_c(z)}{c} \right) &\; \Omega_k < 0
    \end{cases}
    \, .
    \label{eq:dM_ddef}
\end{equation}

\section{Estimating the effect of the inhomogeneities}
\label{sec:VT_estimate}

\subsection{Approximating the efficiency of GW detectors}
\label{sec:VT_estimate:efficiency}

To estimate the effect of the inhomogeneities, we will assume that we can detect all signals with optimum signal to noise ratio (SNR) above a given signal to noise ratio threshold $\rho_\mathrm{thr}$ \footnote{This will be an approximation, since in real gravitational wave data, the measured matched filter SNR will be different from the optimum one due to the noise contribution. Furthermore, real searches for GWs do not just look at the SNR to select candidates~\cite{SPIIR,PyCBC,GstLAL,MBTA}.}

\begin{align}
    \epsilon(\vec{x}, t, \vec{\theta}) & = \Theta(\rho_\mathrm{tot}(\vec{x}, t, \vec{\theta}) - \rho_\mathrm{thr}) \nonumber \\
    & = \begin{cases}
       1 &\; \rho^\mathrm{opt}_\mathrm{tot}(\vec{x}, t, \vec{\theta}) \geq \rho_\mathrm{thr}\\
       0 &\; \rho^\mathrm{opt}_\mathrm{tot}(\vec{x}, t, \vec{\theta}) < \rho_\mathrm{thr}\\
     \end{cases} \, .
    \label{eq:eps_ansatz}    
\end{align}

With the purpose of showcasing the effect that the inhomogeneities can have, we will consider the case of a single mass population of CBCs and taking into account only the inspiral part with the leading order $(l, |m|) = (2 , 2)$ mode. Under these assumptions, the optimum SNR in each interferometer is given by \cite{Maggiore_Vol1}:

\begin{align}
    (\rho^\mathrm{opt}_i)^2 = & \frac{5}{6 \pi^{4/3}} \frac{c^2}{d_L^2} \left( \frac{G \mathcal{M}_c}{c^3} \right)^{5/3} |Q_i(\delta, \alpha,\psi; \iota; t)|^2 \times \nonumber \\ 
    & \times \int_{f_\mathrm{min}}^{f_\mathrm{max}} \d f \frac{f^{-7/3}}{S_i(f)} \, ,
    \label{eq:opt_SNR_Maggiore}
\end{align}

\noindent where $S_i(f)$ is the single sided noise power spectral density (PSD) of the $i$-th detector and $|Q_i(\delta, \alpha,\psi; \iota)|^2$ is a function that captures the dependence on the inclination of the binary $\iota$, the right ascension $\alpha$, declination $\delta$, and polarization angle $\psi$:

\begin{align}
    |Q_i(\delta, \alpha,\psi; \iota; t)|^2 = & \left( F_{+,i}(\delta, \alpha,\psi; t) \frac{1 + \cos^2 \iota}{2} \right)^2 \nonumber \\
    & + \left( F_{\times,i}(\delta, \alpha,\psi; t) \cos\iota\right)^2
    \label{eq:Q_def_Maggiore}
\end{align}

For an interferometer, the antenna patterns as a function of time can be written as~\cite{Jaranowski:1998qm}:

\begin{align}
    F_+(t) & = \sin\zeta\left[a(t)\cos2\psi+b(t)\sin2\psi\right] \, , \label{eq:F+_t}\\
    F_\times(t) & = \sin\zeta\left[b(t)\cos2\psi-a(t)\sin2\psi\right]\ ,  \label{eq:Fx_t}
\end{align}

\noindent where $\zeta$ is the angle between the interferometer arms, and is equal to $90^\circ$ in the case of LIGO, Virgo and Kagra. $a(t)$ and $b(t)$ are the following functions of time:

\begin{align}
a(t) &= \frac{1}{16}\sin2\gamma(3 - \cos2\lambda)(3 - \cos2\delta) \cos[2(\alpha-\phi_r-\Omega_r t)] \nonumber\\
& -\frac{1}{4}\cos2\gamma\sin\lambda(3 - \cos2\delta) \sin[2(\alpha-\phi_r-\Omega_r t)] \nonumber\\
& +\frac{1}{4}\sin2\gamma\sin2\lambda\sin2\delta \cos[\alpha-\phi_r-\Omega_r t] \nonumber\\
&-\frac{1}{2}\cos2\gamma\cos\lambda\sin2\delta \sin[\alpha-\phi_r-\Omega_r t] \nonumber \\
&+\frac{3}{4}\sin2\gamma\cos^2\lambda\cos^2\delta \, , \label{eq:adef}\\
b(t) & = \cos2\gamma\sin\lambda\sin\delta\cos[2(\alpha-\phi_r-\Omega_r t)] \nonumber \\
&+\frac{1}{4}\sin2\gamma(3 - \cos2\lambda)\sin\delta \sin[2(\alpha-\phi_r-\Omega_r t)]\nonumber \\
&+\cos2\gamma\cos\lambda\cos\delta\cos[\alpha-\phi_r-\Omega_r t] \nonumber \\
&+\frac{1}{2}\sin2\gamma\sin2\lambda\cos\delta \sin[\alpha-\phi_r-\Omega_r t] \, , \label{eq:bdef}
\end{align}

\noindent where $\lambda$ is the latitude of the detector’s site, $\Omega_r$ is the rotational angular velocity of the Earth and $\phi_r$ is a phase that defines the position of the Earth~\cite{Jaranowski:1998qm}. $\gamma$ determines the orientation of the detector’s arms with respect to local geographical directions. 

In the case in which we have more than one interferometer, the optimum SNR of each (Eq.~\eqref{eq:opt_SNR_Maggiore}) is added in quadrature, i.e.

\begin{equation}
    (\rho^\mathrm{opt}_\mathrm{tot})^2 = \sum_i (\rho^\mathrm{opt}_i)^2  \, .
    \label{eq:opt_SNR_tot}
\end{equation}

In Eqs.~(\ref{eq:adef},\ref{eq:bdef}) we observe that the dependence of the antenna patterns with the right ascension $\alpha$ and $t$, is periodic and always through the combination $(\alpha-\phi_r-\Omega_r t)$. Therefore, when $T \gg 1\, \mathrm{day}$, the time integral of Eq.~\eqref{eq:eps_avg} will also remove the dependence in $\alpha$ and $\phi_r$, as was argued in Eq.~\eqref{eq:eps_avg_dependence}. 

\begin{figure*}[t!]
\centering
\includegraphics[width=0.495\textwidth]{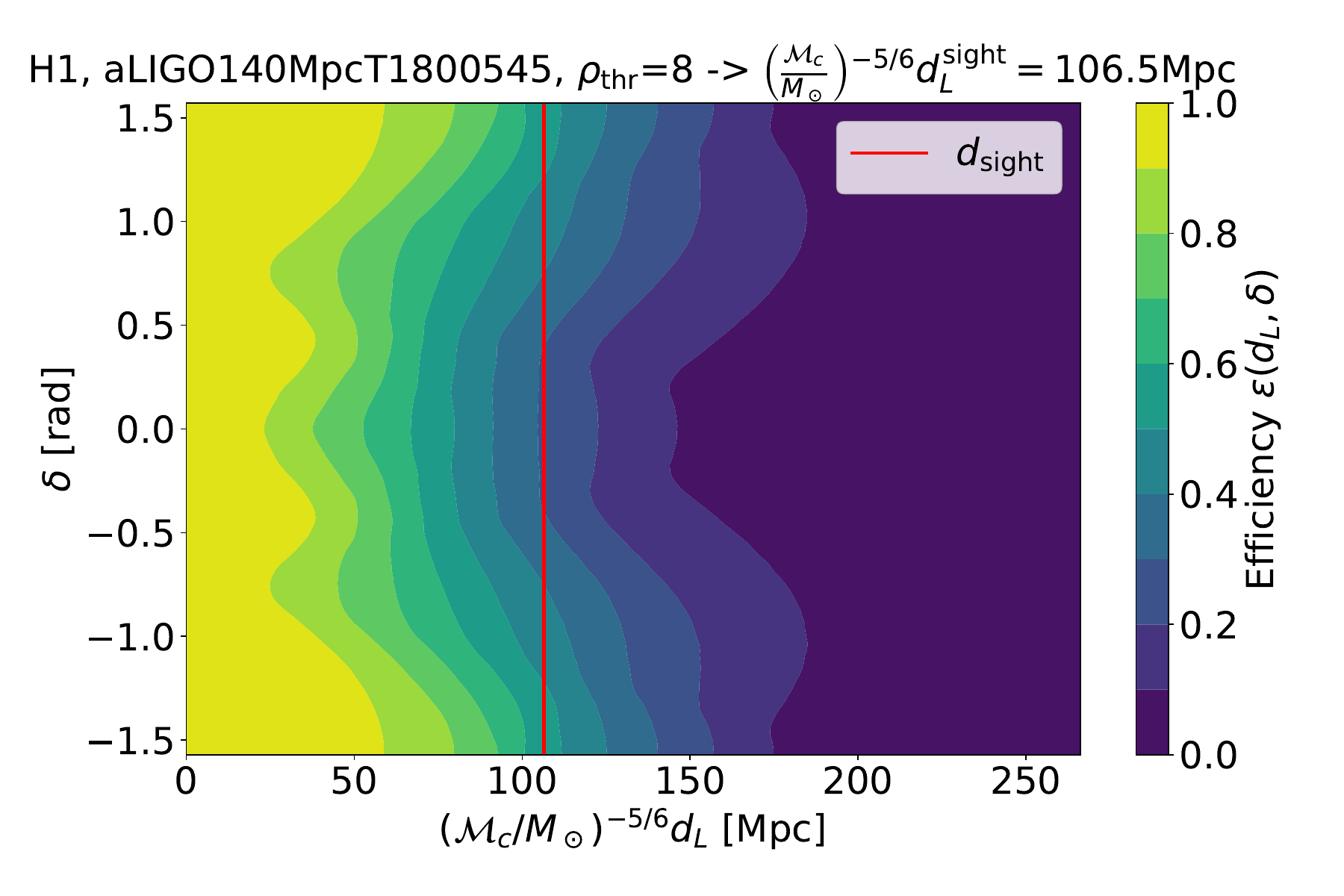}
\includegraphics[width=0.495\textwidth]{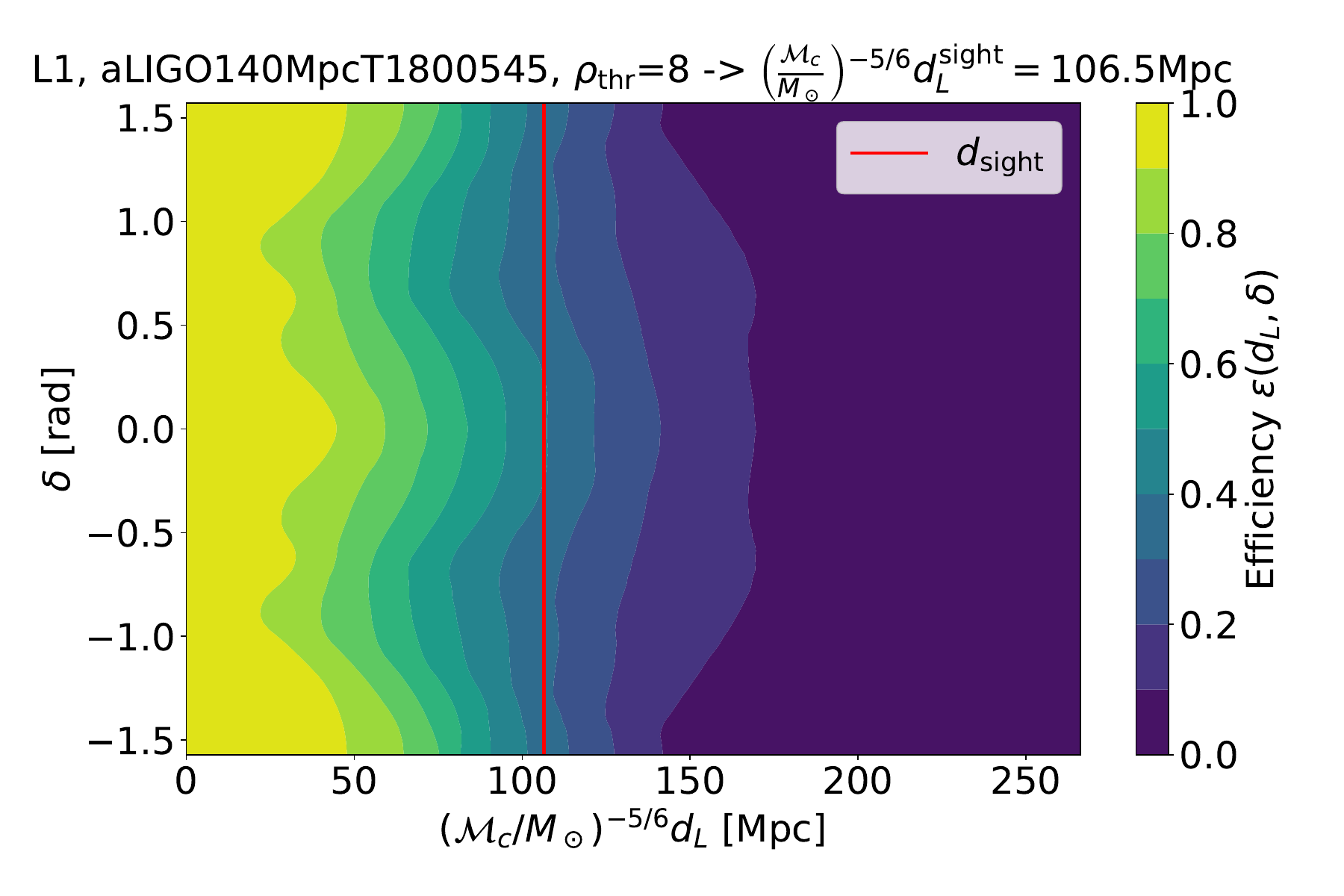}
\includegraphics[width=0.495\textwidth]{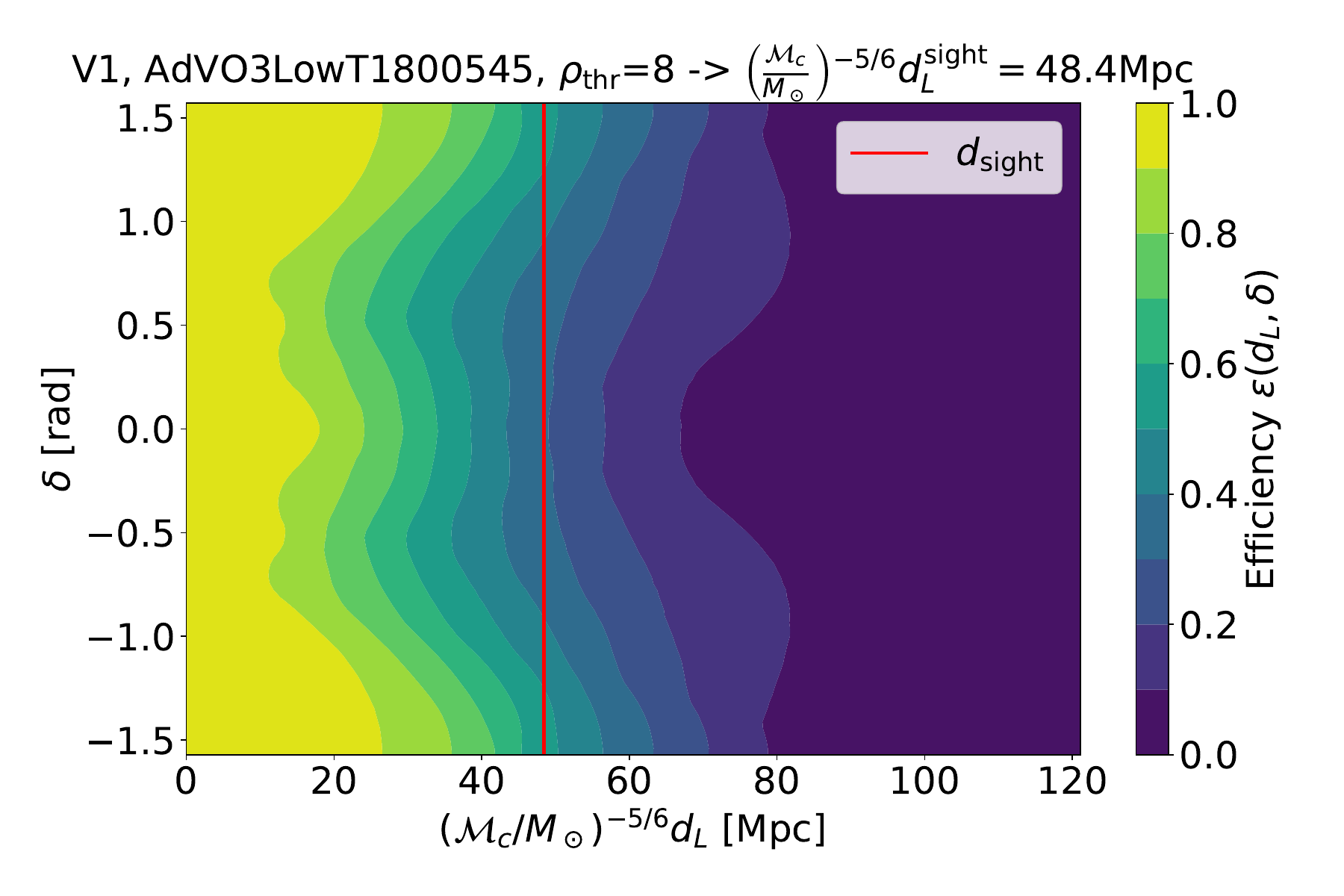}
\includegraphics[width=0.495\textwidth]{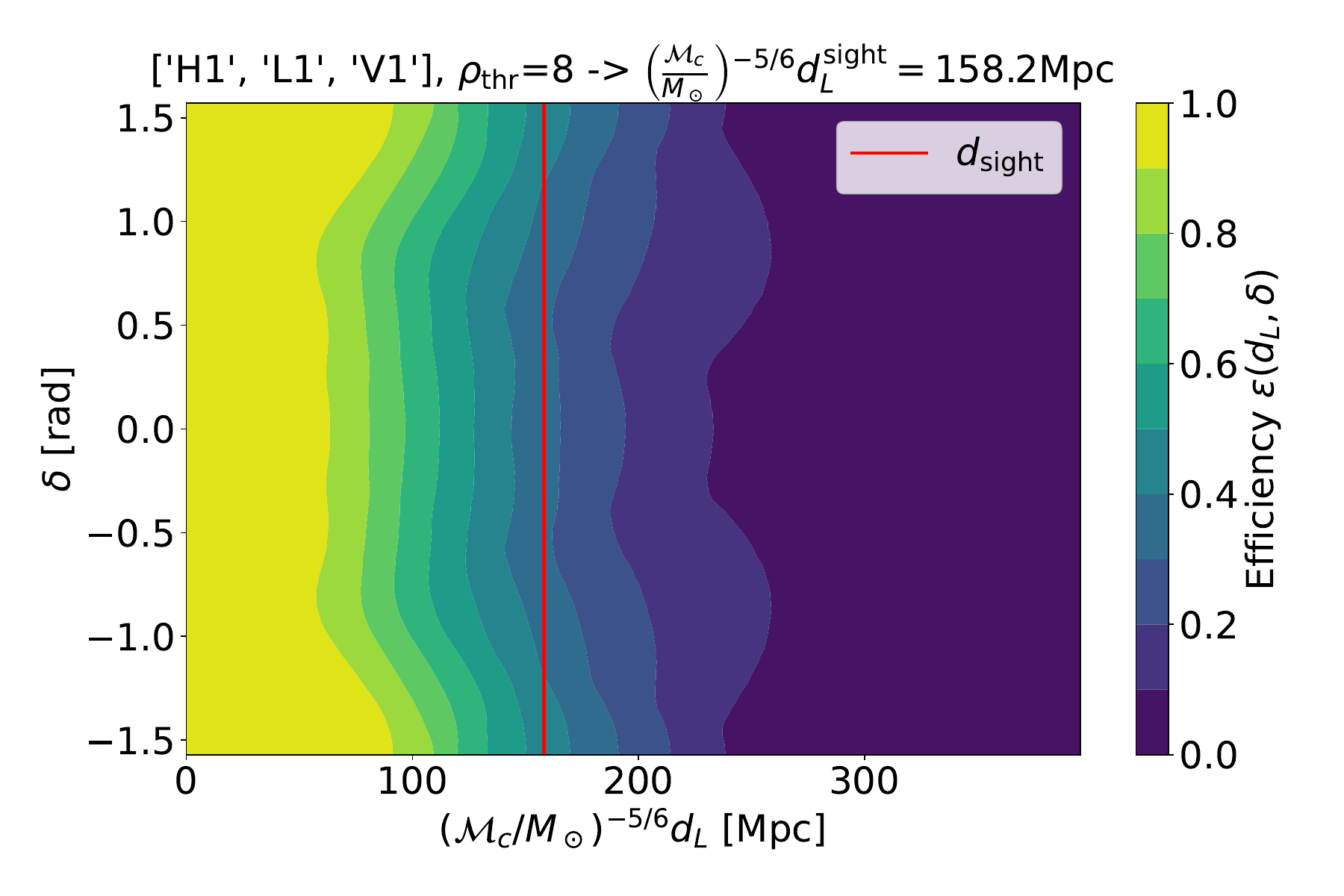}
\caption{Efficiency, or average probability to detect a signal $\epsilon$, defined in Eq.~\eqref{eq:eps_avg_dependence}, as a function of the normalized luminosity distance $d_L$ and the declination $\delta$ for the Hanford (H1), Livingston (L1) and Virgo (V1) interferometers, assuming O3-like sensitivities~\cite{lalsuite,PhysRevD.102.062003, PhysRevLett.123.231108}. We also show the sight distance of each case, computed using Eqs.~(\ref{eq:dsight_Maggiore},\ref{eq:dsight_tot}).  For this plot and all other results in the paper, we assume the optimum SNR threshold of $\rho_\mathrm{thr}=8$.}
\label{fig:efficiencies}
\end{figure*}

In Fig.~\ref{fig:efficiencies} we show the efficiency, or average probability to detect a signal $\epsilon$, defined in Eq.~\eqref{eq:eps_avg_dependence}, as a function of the normalized luminosity distance $d_L$ and the declination $\delta$ for the Hanford (H1), Livingston (L1) and Virgo (V1) interferometers, assuming O3-like sensitivities~\cite{lalsuite,PhysRevD.102.062003,PhysRevLett.123.231108}. We also show the joint sensitivity of the three interferometers at the same time. The efficiencies are simulated integrating Eq.~\eqref{eq:eps_ansatz} with Monte Carlo methods, uniformly sampling over time, polarization $\psi$ and $\cos \iota$. For this plot and all other results in the paper, we assume the optimum SNR threshold of $\rho_\mathrm{thr}=8$~\cite{Morras:2022ysx}.

We observe that $\epsilon(d_L, \delta)$ depends weakly on the declination $\delta$ because interferometers have a very broad angular sensitivity to GWs. The dependence with $\delta$ becomes even weaker when combining the interferometers, since they complement each other. Furthermore, we observe that $\epsilon$ monotonously decreases with distance to the source, since this will make the signals fainter and reduce the SNR. The way that $\epsilon(d_L, \delta)$ decays with distance is determined by the sensitivity of the GW detectors involved. To quantify the sensitivity one usually uses the sight distance~\cite{Maggiore_Vol1}, which is computed by averaging $(\rho^\mathrm{opt}_i)^2$ (Eq.~\eqref{eq:opt_SNR_Maggiore}) over angles and equating it to the SNR threshold. Using that $\langle |Q|^2 \rangle = 2/5$, we have:

\begin{equation}
    d_{L,i}^\mathrm{sight} = \frac{1}{\rho_\mathrm{thr}} \frac{2}{5} \sqrt{\frac{5}{6}} \frac{c}{\pi^{2/3}} \left( \frac{G \mathcal{M}_c}{c^3} \right)^{5/6} \sqrt{\int_{f_\mathrm{min}}^{f_\mathrm{max}} \d f \frac{f^{-7/3}}{S_i(f)}} \, .
    \label{eq:dsight_Maggiore}
\end{equation}

When we have multiple detectors, to be consistent with Eq.~\eqref{eq:opt_SNR_tot}, we define the sight distance of the network as the result of adding the single detector sight distances in quadrature:

\begin{equation}
    (d_{L,\mathrm{tot}}^\mathrm{sight})^2 = \sum_i (d_{L,i}^\mathrm{sight})^2\, .
    \label{eq:dsight_tot}
\end{equation}

To see the dependance of the efficiency as a function of distance, we can define the declination averaged efficiency as
\begin{equation}
    \Bar{\epsilon}(d_L) = \int_{-1}^{1} \frac{\d\sin{(\delta)}}{2} \, \epsilon(d_L,\delta) \, .
    \label{eq:dec_avg_eff}
\end{equation}

\begin{figure}[t!]
\centering  
\includegraphics[width=0.495\textwidth]{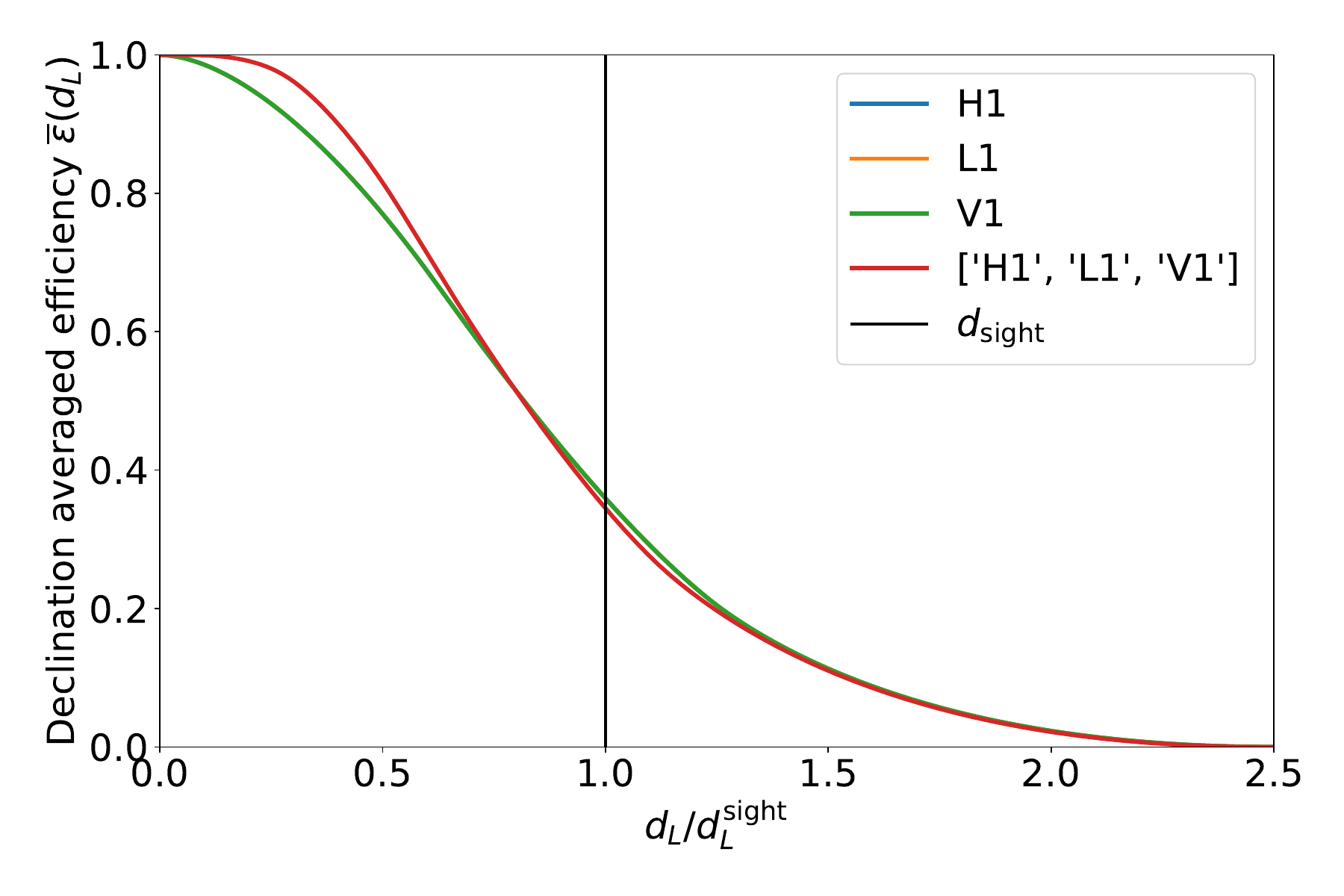}
\caption{Declination averaged efficiency $\Bar{\epsilon}(d_L)$ as defined in Eq.~\eqref{eq:dec_avg_eff} for the same cases as in Fig.~\ref{fig:efficiencies}. We normalize the luminosity distance by the sight distance of each case as defined in Eqs.~(\ref{eq:dsight_Maggiore}, \ref{eq:dsight_tot}). Note that the lines for ``H1'', ``L1'' and ``V1'' overlap, since the single detector declination averaged efficiencies are all the same.}
\label{fig:efficiencies_dec_avg}
\end{figure}

In Fig.~\ref{fig:efficiencies_dec_avg} we show this declination averaged efficiency $\Bar{\epsilon}(d_L)$ for the same cases as in Fig.~\ref{fig:efficiencies}. We observe that in all single interferometer cases, $\Bar{\epsilon}(d_L)$ behaves in the same way, having $100\%$ efficiency at $d_L = 0$ and slowly losing efficiency as the CBCs are placed farther until we reach $0\%$ efficiency at $(5/2)d_{L}^\mathrm{sight}$. For the case of the three detector network, the shape of $\Bar{\epsilon}(d_L)$ is slightly different, having a bit of a sharper drop-off.

\subsection{Inhomogeneities in the local Universe}
\label{sec:VT_estimate:inhomo}

For the purposes of showcasing the effects of the inhomogeneities in the measured rate of CBCs we will make the assumption that the rate is proportional to the matter density $\rho(\vec{x})$

\begin{equation}
    R(\vec{x}) \propto \rho(\vec{x}) \, .
    \label{eq:R_propto_rho}
\end{equation}

We do not expect this assumption to be true on very small scales, since the processes that generate binaries of compact objects that merge in the age of the Universe will strongly depend on the properties of their local environment. However, in this section, we are going to explore merger rate inhomogeneities on large scales, of the order of megaparsecs (Mpc), over which we can assume that regions generating CBCs represent a similar fraction of matter, and therefore that the merger rate density is proportional to the matter density. 

Furthermore, since we will be exploring small redshifts ($z \lesssim 0.08$), and to avoid confounding factors in showcasing the effects of the inhomogeneities, we will ignore the evolution of the merger rates with cosmic time, i.e. we assume that $\overline{R}(z) \approx \overline{R}_0$. Under these conditions, the event rate inhomogeneities will be equal to the matter over-density

\begin{equation}
    \delta_R(\vec{x}) = \delta_M(\vec{x}) = \frac{\rho(\vec{x}) - \overline{\rho}}{\overline{\rho}}\, .
    \label{eq:dR_equal_drho}
\end{equation}

\begin{figure}[t!]
\centering  
\includegraphics[width=0.495\textwidth]{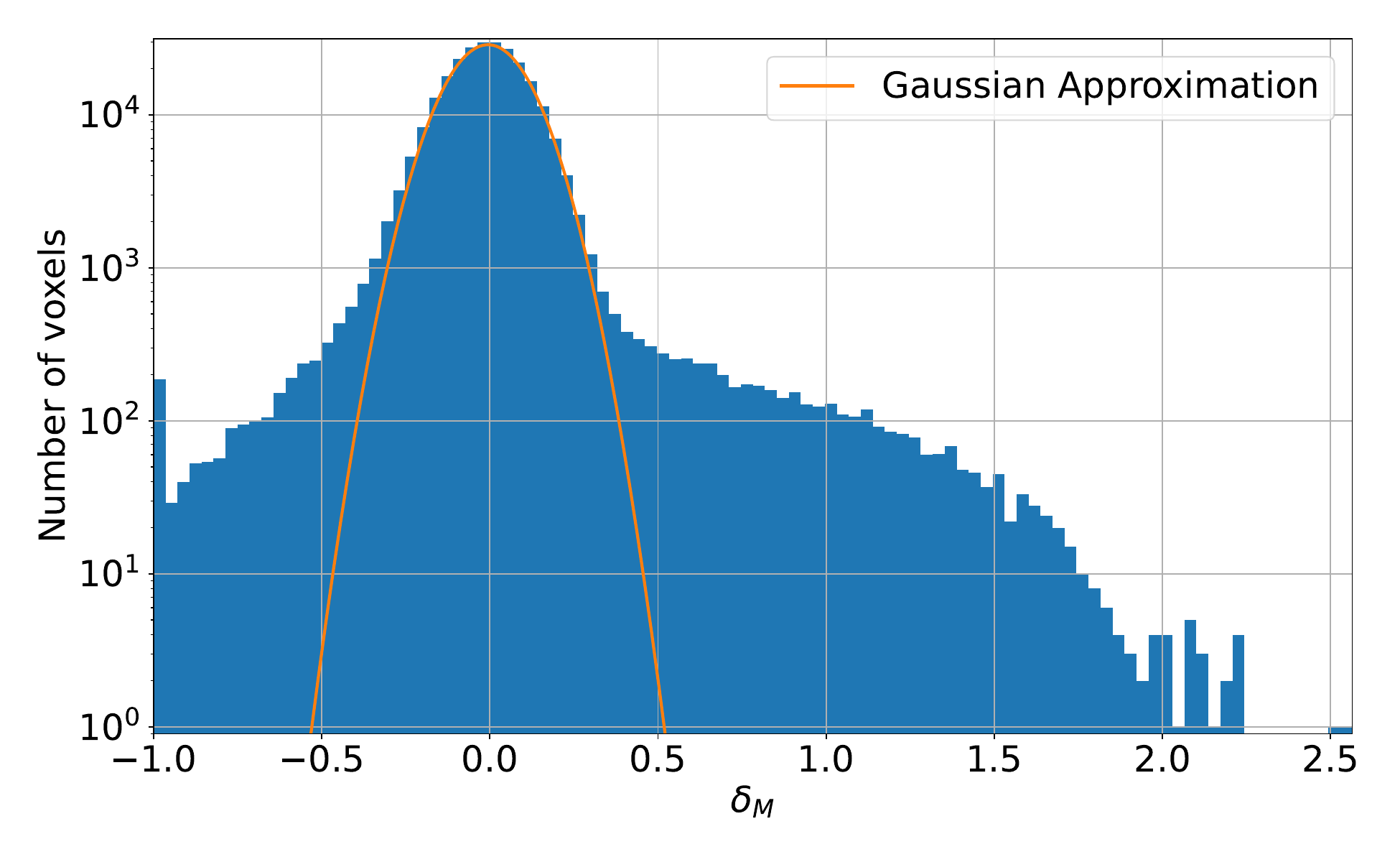}
\caption{Histogram of the value of $\delta_M$ in the voxels of Ref.~\cite{Courtois:2022mxo}, reconstructed from the grouped CF4. The total number of voxels is $64^3 = 262144$. We also show a Gaussian approximation, fitted in the range $\delta_M \in [-0.1, 0.1]$.}
\label{fig:delta_histo_2211.16390}
\end{figure}

These matter over-density have been well measured in the local Universe using the peculiar (i.e. gravitational) velocities of galaxies~\cite{Courtois:2022mxo} from the CosmicFlows (CF) catalogs~\cite{Tully:2016ppz,Tully:2022rbj}. In particular we take the $\delta_M(\vec{x})$ data published in Ref.~\cite{Courtois:2022mxo}, reconstructed from the grouped CF4, which reaches a redshift of $z=0.08$, i.e. a distance of $d \sim 340\mathrm{Mpc}/h_{70}$ having $64^3$ voxels\footnote{A voxel represents a value on a regular grid in three-dimensional space. It is the gerneralization to 3D of a 2D pixel.}. In Fig.~\ref{fig:delta_histo_2211.16390} we show the distribution of the over-density $\delta_M$ in each of these voxels. We observe that while the distribution is peaked at $\delta_M = 0$, being well approximated by a Gaussian function in the $|\delta_M | \ll 1$ regime, it presents very large non-Gaussian tails which are a direct consequence of the very non-linear structure present in our Universe, formed by large voids ($\delta_M \sim -1$) and over-dense regions ($\delta_M \gtrsim 1$). 

To give a first estimate of how much these inhomogeneities will affect the merger rate density estimation, we temporarily assume that the probability to detect a signal does not depend on declination (i.e. $\epsilon(z,\delta) = \epsilon(z)$). Then, the $\langle VT \rangle$ could be computed as:

\begin{equation}
    \langle V T \rangle = T \int_0^{z_\mathrm{max}} \d z \frac{dV_c}{dz} P(z) \epsilon(z) \frac{1}{1 + z} \, ,
    \label{eq:VT_isotrop}
\end{equation}

\noindent where we have defined the angle averaged rate enhancement $P(z)$ as

\begin{equation}
    P(z) = \frac{1}{4\pi} \int_0^{2 \pi} \d\phi \int_0^{\pi} \sin{\theta} \d\theta \, (1 + \delta_M(z,\theta,\phi)) \, .
    \label{eq:Pz_def}
\end{equation}

\begin{figure}[h!]
\centering  
\includegraphics[width=0.495\textwidth]{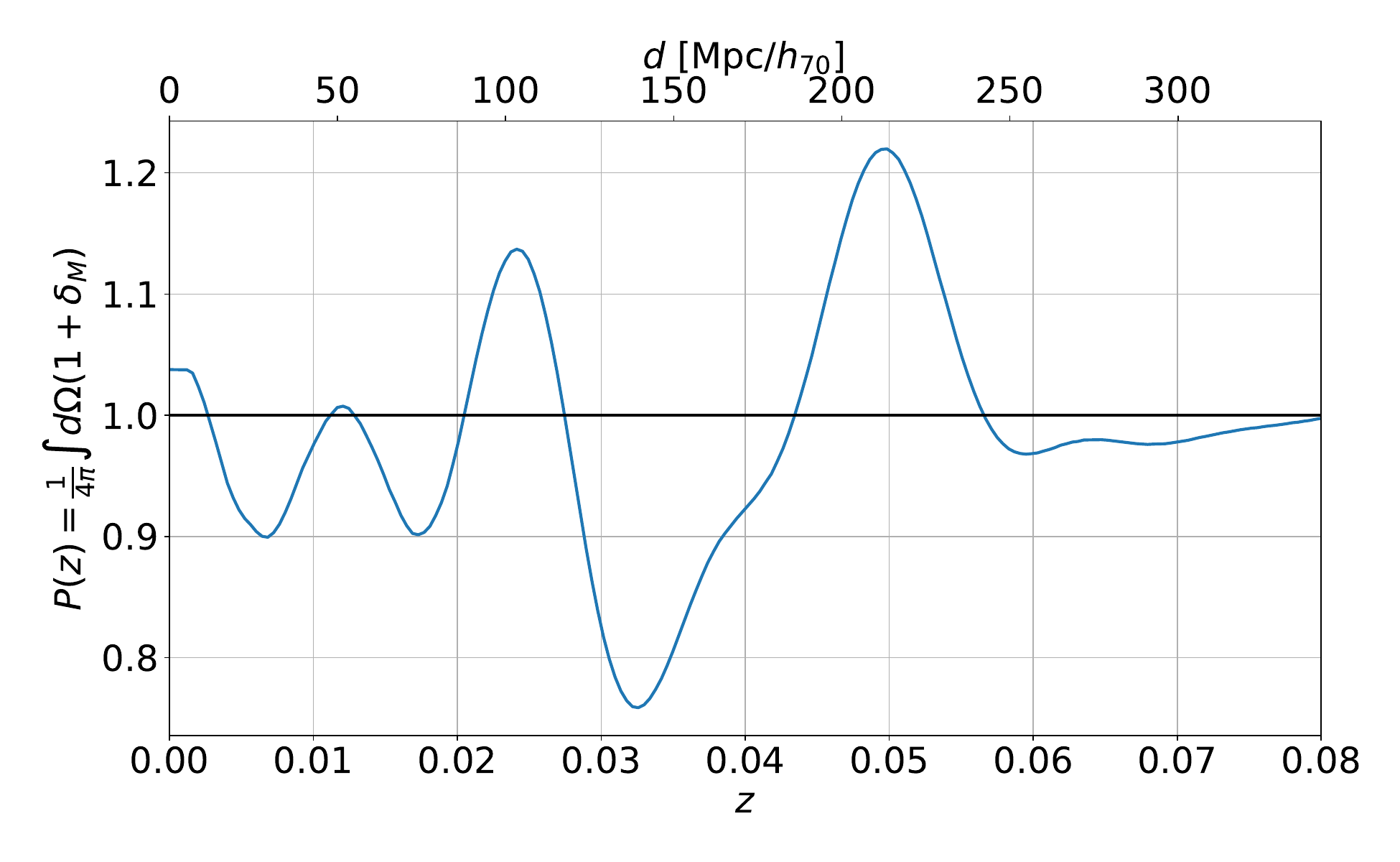}
\caption{Angle averaged rate enhancement, as defined in Eq.~\eqref{eq:Pz_def} and computed using the $\delta_M(z,\theta,\phi)$ of Ref.~\cite{Courtois:2022mxo}.}
\label{fig:Pz_2211.16390}
\end{figure}

In Fig.~\ref{fig:Pz_2211.16390} we show this quantity as a function of redshift. We observe that deviations from 1 in the angle averaged rate enhancement $P(z)$ can be very large, of more than 20\%. At large $z$ we expect $P(z) \to 1$, since we will be averaging the over-densities over a larger surface in the integral of Eq.~\eqref{eq:Pz_def}. This is what is observed in Fig.~\ref{fig:Pz_2211.16390}. Note however, that there is a very large suppression at $z \sim 0.03$ ($d \sim 140~\mathrm{Mpc}/h_{70}$) followed by a large enhancement at $z \sim 0.05$ ($d \sim 210~\mathrm{Mpc}/h_{70}$), which are relatively large distances. These are caused by the Sloan Great Wall~\cite{Gott:2003pf}, a huge structure in the local Universe, which spans about $400~\mathrm{Mpc}/h_{70}$. The Sloan Great Wall is preceded by a correspondingly huge void, as can be seen in Fig.~\ref{fig:Pz_2211.16390}.

\subsection{Effect of inhomogeneities in current GW detectors}
\label{sec:VT_estimate:estimate}

In this section we will put together the results of Sec.~\ref{sec:VT_estimate:efficiency} and Sec.~\ref{sec:VT_estimate:inhomo} to give an estimate of the effect of inhomogeneities in current GW detectors.

Since we will consider our Universe to be flat ($\Omega_k \sim 0$~\cite{Planck:2018vyg}) and the inhomogeneity data that we are using~\cite{Courtois:2022mxo} contains redshifts $z \ll 1$, in Eq.~\eqref{eq:dVc_dc} we have:

\begin{equation}
    d_c(z) = d_M(z) \approx \frac{c}{H_0} z \left( 1 - \frac{3}{4} \Omega_m z + O(z^2) \right)
    \label{eq:dVc_approx}
\end{equation}

\noindent where $\Omega_m = 0.315 \pm 0.007$~\cite{Planck:2018vyg}, and the integral of Eq.~\eqref{eq:VT_eps_avg} can be approximated as:

\begin{align}
    \langle V T \rangle  \approx  T  \left( \frac{c}{H_0} \right)^3 \int_\Omega  \d\Omega  \int_0^{z_\mathrm{max}} \d z \, z^2 (1 + \delta_R(\vec{x})) \epsilon(z, \delta) \times \nonumber \\
    \times \left[1 - (3 \Omega_m + 1) z +O(z^2) \right] \, .
    \label{eq:VT_integral_approx}
\end{align}

For $z<0.08$, the value of the $O(z^2)$ term we are ignoring is smaller than $1.5\%$. 

To estimate the effect that inhomogeneities have on the $\langle V T \rangle$ of current GW detectors, we will use Eq.~\eqref{eq:VT_integral_approx} to compute the ratio between the $\langle V T \rangle_\mathrm{inhomo}$, obtained setting $\delta_R$ to the matter inhomogeneities $\delta_M$ of Ref.~\cite{Courtois:2022mxo}, and $\langle V T \rangle_\mathrm{homo}$ assuming homogeneity (i.e. $\delta_R=0$). When computing the ratio $\langle V T \rangle_\mathrm{inhomo}/\langle V T \rangle_\mathrm{homo}$ we expect effects of the dependence of $\overline{R}(z)$ with cosmic time, which is common in both cases, to largely cancel out.

\begin{figure}[h!]
\centering  
\includegraphics[width=0.495\textwidth]{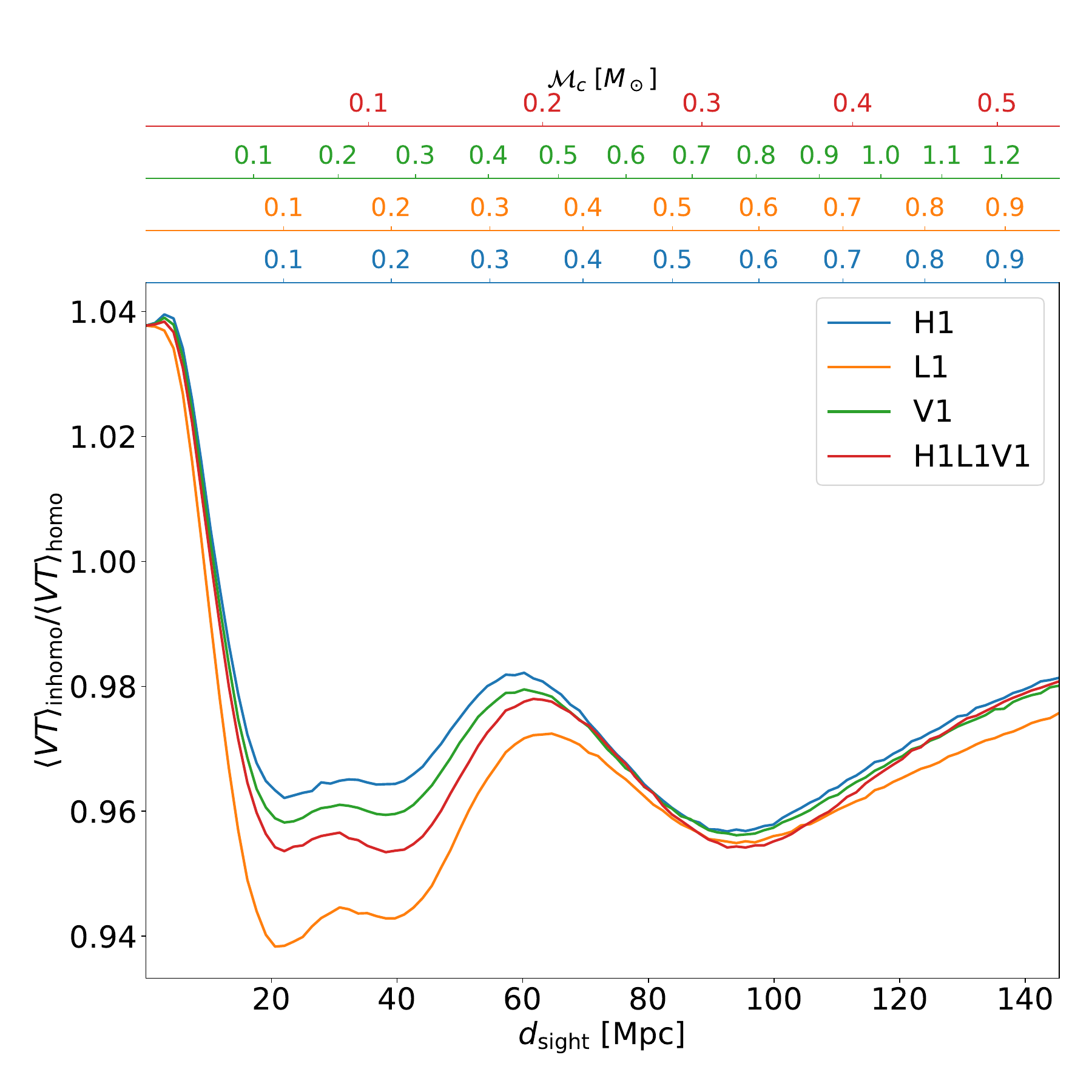}
\caption{Ratio between $\langle V T \rangle\mathrm{s}$ computed using Eq.~\ref{eq:VT_integral_approx} setting $\delta_R(\vec{x}) = \delta_M(\vec{x})$ from Ref.~\cite{Courtois:2022mxo} ($\langle V T \rangle_\mathrm{inhomo}$) and setting $\delta_R(\vec{x}) = 0$ ($\langle V T \rangle_\mathrm{homo}$). The maximum sight distance shown of $d_\mathrm{sight} \sim 145~\mathrm{Mpc} = (2/5) d_L(z=0.08)$ is set to avoid the tails of the efficiencies (see Fig.~\ref{fig:efficiencies_dec_avg}) exiting the available data.}
\label{fig:enh_VT_dsight_ifos}
\end{figure}

The ratio $\langle V T \rangle_\mathrm{inhomo}/\langle V T \rangle_\mathrm{homo}$ is shown as a function of the sight distance for the different detectors in Fig.~\ref{fig:enh_VT_dsight_ifos}. We observe that this ratio has deviations of at most $6\%$ from unity, much smaller than the $O(20\%)$ deviations that were observed in the angle averaged rate enhancement $P(z)$ of Fig.~\ref{fig:Pz_2211.16390}. The reason the effect is suppressed is mainly because the efficiency is very extended in redshift (as can be seen in Fig.~\ref{fig:efficiencies_dec_avg}), averaging the effect of the inhomogeneities over different redshifts. The amplitude of the effect is very similar between interferometers, because as we saw in Fig.~\ref{fig:efficiencies} the efficiency depends weakly on the inclination, however there are some small differences due to their different orientations with respect to the structure in the local Universe. 

Furthermore, looking at the top axes of Fig.~\ref{fig:enh_VT_dsight_ifos} we can see that different sight distances will correspond to different CBC chirp masses (assuming O3-like sensitivities). As can be seen from Eq.~\eqref{eq:dsight_Maggiore}, the more sensitive a detector is (smaller $S_n$), the larger the sight distance at a given mass. Even for O3-like sensitivities, the effects of the inhomogeneities on the $\langle VT \rangle$ estimation, which is relevant for $d_L \lesssim 140\mathrm{Mpc}$ will correspond to $\mathcal{M}_c \lesssim 1 M_\odot$ in the single detector case and $\mathcal{M}_c \lesssim 0.5 M_\odot$ in the case in which we consider the full H1-L1-V1 network. Therefore these results will be most relevant for the computation of the $\langle VT \rangle$ in searches of sub-solar mass (SSM) black holes~\cite{LIGOScientific:2022hai}, which are used to put constraints on the merger rates of such objects and in turn to constrain models such as primordial black holes~\cite{Carr:2023tpt, Escriva:2022duf} and dark matter black holes~\cite{Shandera:2018xkn}.

For Cosmic Explorer~\cite{Evans:2021gyd} and Einstein Telescope~\cite{Maggiore_2020,Branchesi:2023mws} we expect similar results to that of Fig.~\ref{fig:enh_VT_dsight_ifos}. However, due to the much higher sensitivities of these interferometers, we expect the luminosity distances to correspond to much smaller chirp masses, and therefore the effect will only be relevant for even smaller black holes. The durations of the signals from these tiny black holes inside the detector sensitivity band will be much longer and they would probably have to be looked for with continuous wave methods~\cite{Miller:2020kmv,Alestas:2024ubs}; instead of the matched filtering approach we have based our modeling of $\epsilon(z,\delta)$ on.

\section{Conclusions}
\label{sec:conclusions}

In this paper we have discussed the effect of inhomogeneities on the estimated merger rates of GW events, which can bias the measured local rate of GW events due to our particular location in the Universe. In Sec.~\ref{sec:VT} we show that the $\langle VT \rangle$, the quantity used to reconstruct the merger rates in GW data analysis, is given by the observation time $T$ multiplied by the integral over comoving volume of the population and time averaged efficiency $\epsilon(\vec{x})$, weighted by the normalized merger rate density $(1+\delta_R(\Vec{x}))$.

We have then studied the effect of the inhomogeneous matter distribution of our local Universe, obtained from CosmicFlows-4~\cite{Courtois:2022mxo}, on the estimated rate of gravitational waves events in the LIGO/Virgo interferometers, assuming that the spatial distribution of events is proportional to the matter distribution. In Fig.~\ref{fig:efficiencies} we observe that the effect of the angular inhomogeneities will be smoothed out by the low angular resolution of the interferometers, coupled with the rotation of the earth, which sweeps away any dependence of the average detection efficiency with the right ascension $\alpha$. 

Furthermore, even though, as seen in Fig.~\ref{fig:Pz_2211.16390}, the redshift dependence of the local matter distribution has large variations (of the order of 20\%) in the density contrast along the line of sight (averaged over spherical shells), the smooth dependence with redshift of the efficiency function (seen in Fig.~\ref{fig:efficiencies_dec_avg}), again averages out the inhomogeneities in redshift. In Fig.~\ref{fig:enh_VT_dsight_ifos} we show that the local inhomogeneities give at most a 6\% effect on the $\langle V T \rangle$ when compared with the homogeneous assumption, having a dependence with the sight distance, which will correspond with different chirp masses depending on the detector sensitivity. There is a trend; for distances below $\sim$10 Mpc, the ratio is always larger for the inhomogeneous distribution, while for distances from $\sim$10 to $\sim$200 Mpc it is always smaller. With sufficient statistics (i.e. hundreds of events in the nearby Universe) we may be able to confirm whether GW events are correlated with the inhomogeneous matter distribution, as expected, or if they are uniformly distributed across the sky.

Extending the analysis to future detectors like the Einstein Telescope and Cosmic Explorer, with significantly better sensitivities, we can expect to be able to detect BBH mergers with smaller chirp masses at larger distances, and therefore be sensitive to the local matter inhomogeneities for smaller masses. One could then test the hypothesis of proportionality between BBH event rates and matter density. Moreover, for a large population of binary neutron star mergers, this may place a significant constraint on the local matter distribution~\cite{Essick:2022slj,Vitale:2022pmu}.

Note that this analysis extends from the local group of galaxies to the cosmological homogeneity scale. However, for other type of events, like supernovae explosions~\cite{Szczepanczyk:2021bka} or close hyperbolic encounters~\cite{Garcia-Bellido:2017qal,Garcia-Bellido:2017knh,Bini:2023gaj}, the relevant inhomogeneities are those within our own galaxy. This requires a detailed 3D reconstruction of the Milky Way which may be possible in the near future with surveys like Gaia~\cite{TheGaiamission}. We leave this analysis for a future work.

\section*{Acknowledgements}

The authors thank Maya Fishbach and Shanika Galaudage for their helpful comments and discussions as internal reviewers of this paper in LIGO and Virgo respectively.
The authors acknowledge support from the research project  PID2021-123012NB-C43 and the Spanish Research Agency (Agencia Estatal de Investigaci\'on) through the Grant IFT Centro de Excelencia Severo Ochoa No CEX2020-001007-S, funded by MCIN/AEI/10.13039/501100011033. GM acknowledges support from the Ministerio de Universidades through Grant No. FPU20/02857.
This material is based upon work supported by NSF's LIGO Laboratory which is a major facility fully funded by the National Science Foundation.

\bibliography{Refs}

\end{document}